\documentclass[10pt,A4paper,conference]{IEEEtran}
\usepackage{amsthm}
\usepackage{amsmath}
\usepackage{amsfonts}
\usepackage{graphicx}
\usepackage{latexsym}
\usepackage{amssymb}
\usepackage{stmaryrd}
\usepackage{amscd}
\usepackage[small]{caption}

\newcommand{\deff}{\mbox{$\stackrel{\rm def}{=}$}}

\newcommand{\sbinom}[2]{\left[ \begin{array}{c} #1 \\ #2 \end{array} \right] }

\newcommand{\field}[1]{\mathbb{#1}}

\newcommand{\Z}{\field{Z}}

\newcommand{\cC}{{\cal C}}

\newcommand{\cK}{{\cal K}}

\newcommand{\cR}{{\cal R}}

\newcommand{\cM}{{\cal M}}

\newcommand{\sP}{\field{P}}

\newcommand{\sG}{\field{G}}

\DeclareMathAlphabet{\mathbfsl}{OT1}{cmr}{bx}{it}
\newcommand{\uuu}{\kern-1pt\mathbfsl{u}\kern-0.5pt}
\newcommand{\vvv}{\kern-1pt\mathbfsl{v}\kern-0.5pt}

\newcommand{\myboxplus}{\kern1pt\mbox{\small$\boxplus$}}

\makeatletter \DeclareRobustCommand{\sbinom}{\genfrac[]\z@{}}
\makeatother
\newcommand{\G}[2]{\sbinom{{#1}\kern-1pt}{{#2}\kern-1pt}}
\newcommand{\Gq}[2]{\sbinom{{#1}\kern-0.25pt}{{#2}\kern-0.25pt}}

\newcommand{\Ps}{\smash{{\sP\kern-2.0pt}_q\kern-0.5pt(n)}}
\newcommand{\sPs}{\smash{{\sP\kern-1.5pt}_q(n)}}
\newcommand{\Ptwo}{\smash{{\sP\kern-2.0pt}_2\kern-0.5pt(n)}}
\newcommand{\Ptwom}{\smash{{\sP\kern-2.0pt}_2\kern-0.5pt(m)}}
\newcommand{\Ptwonm}{\smash{{\sP\kern-2.0pt}_2\kern-0.5pt(n+m)}}
\newcommand{\Ptwoa}{\smash{{\sP\kern-2.0pt}_2\kern-0.5pt(1)}}
\newcommand{\Ptwob}{\smash{{\sP\kern-2.0pt}_2\kern-0.5pt(2)}}
\newcommand{\Ptwoc}{\smash{{\sP\kern-2.0pt}_2\kern-0.5pt(3)}}
\newcommand{\Ptwod}{\smash{{\sP\kern-2.0pt}_2\kern-0.5pt(4)}}
\newcommand{\Ptwoe}{\smash{{\sP\kern-2.0pt}_2\kern-0.5pt(5)}}
\newcommand{\Ptwof}{\smash{{\sP\kern-2.0pt}_2\kern-0.5pt(6)}}
\newcommand{\Ptwokm}{\smash{{\sP\kern-2.0pt}_2\kern-0.5pt(2k-1)}}
\newcommand{\Pone}{\smash{{\sP\kern-2.5pt}_2\kern-0.5pt(n{-}1)}}

\newcommand{\Gr}{\smash{{\sG\kern-1.5pt}_q\kern-0.5pt(n,k)}}
\newcommand{\Gi}{\smash{{\sG\kern-1.5pt}_q\kern-0.5pt(n,i)}}
\newcommand{\Gj}{\smash{{\sG\kern-1.5pt}_q\kern-0.5pt(n,j)}}
\newcommand{\Grmk}{\smash{{\sG\kern-1.5pt}_q\kern-0.5pt(n,n-k)}}
\newcommand{\Grdk}{\smash{{\sG\kern-1.5pt}_q\kern-0.5pt(2k,k)}}
\newcommand{\Grekappa}{\smash{{\sG\kern-1.5pt}_q\kern-0.5pt(n,e+1-\kappa)}}
\newcommand{\Grtwoekappa}{\smash{{\sG\kern-1.5pt}_q\kern-0.5pt(n,2e+1-\kappa)}}
\newcommand{\Gremkappa}{\smash{{\sG\kern-1.5pt}_q\kern-0.5pt(n,e-\kappa)}}
\newcommand{\Gn}{\smash{{\sG\kern-1.5pt}_2\kern-0.5pt(n,n{-}1)}}
\newcommand{\Gnq}{\smash{{\sG\kern-1.5pt}_q\kern-0.5pt(n,n{-}1)}}
\newcommand{\Gone}{\smash{{\sG\kern-1.5pt}_2\kern-0.5pt(n,1)}}
\newcommand{\Gqone}{\smash{{\sG\kern-1.5pt}_q\kern-0.5pt(n,1)}}
\newcommand{\GTwo}{\smash{{\sG\kern-1.5pt}_2\kern-0.5pt(n,k)}}
\newcommand{\GTwonk}[2]{{\smash{{\sG\kern-1.5pt}_2\kern-0.5pt({#1},{#2})}}}
\newcommand{\Gnk}{\smash{{\sG\kern-1.5pt}_2\kern-0.5pt(n,n{-}k)}}
\newcommand{\Greone}{\smash{{\sG\kern-1.5pt}_q\kern-0.5pt(n,e{+}1)}}
\newcommand{\Gretwo}{\smash{{\sG\kern-1.5pt}_q\kern-0.5pt(n,e{+}2)}}

\newcommand{\be}[1]{\begin{equation}\label{#1}}
\newcommand{\ee}{\end{equation}}

\newcommand{\Cref}[1]{Co\-rol\-la\-ry\,\ref{#1}}

\newtheorem{theorem}{Theorem}
\newtheorem{lemma}{Lemma}

\newtheorem{cor}{Corollary}

\newtheorem{example}{Example}

\begin{document}

\vspace{-0.5cm}

\title{Systematic Codes for Rank Modulation\vspace{-1.0ex}}

\author{\textbf{Sarit Buzaglo}\IEEEauthorrefmark{1}, \IEEEauthorblockN{\textbf{Eitan Yaakobi}\IEEEauthorrefmark{1}, \textbf{Tuvi Etzion}\IEEEauthorrefmark{1}, and \textbf{Jehoshua Bruck}\IEEEauthorrefmark{2}}
\IEEEauthorblockA{\IEEEauthorrefmark{1}Computer Science Department, Technion -- Israel Institute of Technology, Haifa 32000, Israel \\}
\IEEEauthorblockA{\IEEEauthorrefmark{2}Electrical Engineering Department, California Institute of Technology, Pasadena, CA 91125, U.S.A\\}
{\it  \{sarahb, yaakobi, etzion\}@cs.technion.ac.il}, {\it bruck@caltech.edu}\vspace{-0.13ex}}

\maketitle

\begin{abstract}
The goal of this paper is to construct systematic error-correcting
codes for permutations and multi-permutations in the Kendall's $\tau$-metric.
These codes are important in
new applications such as rank modulation for flash memories.
The construction is based on error-correcting codes for multi-permutations
and a partition of the set of permutations into error-correcting codes.
For a given large enough number of information symbols $k$,
and for any integer~$t$, we present a construction for
${(k+r,k)}$ systematic $t\text{-}$error-correcting codes, for permutations
from~$S_{k+r}$, with
less redundancy symbols than the number of redundancy
symbols in the codes of the known constructions.
In particular, for a given $t$ and for sufficiently large $k$ we
can obtain $r=t+1$.
The same construction is also applied to obtain related systematic
error-correcting codes for multi-permutations.
\end{abstract}

\section{Introduction}

Flash memory is one of the most widely used non-volatile technology.
In flash memories, cells usually represent multiple
levels, which correspond to the amount of electrons trapped in each cell.
Currently, one of the main challenges in flash memory cells is to
program each cell exactly to its designated level. In order to overcome this difficulty,
the novel framework of \emph{rank modulation codes} was introduced in~\cite{JMSB09}.
In this setup, the information is carried by the relative values between the cells
rather than by their absolute levels. Thus, every group of cells
induces a permutation, which is derived by the ranking of the level of each
cell in the group. There are several works which study the correction of errors
under the setup of permutations for the rank modulation scheme; see e.g.~\cite{BM10,JSB10,TaSc10,TaSc12,ZJB12,ZMJB13}.
In all these works $t$-error-correcting codes were considered for the set $S_n$, which consists
of all permutations on $n$ elements, endowed with either
the Kendall's $\tau$-metric or the infinity metric.
Recently, to improve the number of rewrites,
the model of rank modulation was extended such that multiple cells can
share the same ranking~\cite{EJB12,EYJB13}. Thus, the cells no longer
determine permutations but rather multi-permutations, which are also known as permutations
with repetitions. Error-correcting
codes for multi-permutations subject to the Kendall's $\tau$-metric were presented in~\cite{SGD12}
and also studied in~\cite{EYEB13}.

The main goal of this paper is to construct \emph{systematic}
error-correcting codes for permutations.
This concept for permutations was proposed in~\cite{ZJB12,ZMJB13}.
In a systematic code $\cC$ for permutations in $S_n$ we have $k!$ codewords.
Each permutation of $S_k$ (on a given set of specific $k$ symbols) is
a sub-permutation of exactly one codeword of $\cC$.
In this paper we improve on some of the results in~\cite{ZJB12,ZMJB13}.
Our construction of systematic error-correcting codes
for permutations is based on two ingredients.
The first is a partition of~$S_k$ into $t$-error-correcting codes.
The second is a code~$\cC_r$ for multi-permutations from the multi-set
$\{ 0^k, k+1,\ldots , k+r \}$ with minimum Kendall's $\tau$-distance~$2t$, whose
size is the number of parts in the partition.
Each code from the partition of~$S_k$ will be substituted into a different
codeword of $\cC_r$. This construction will be generalized to systematic codes
for multi-permutations.

The rest of this work is organized as follows. Our
construction is heavily based on error-correcting codes for
multi-permutations. Hence, in
Section~\ref{sec:basic} we define the basic concepts for multi-permutations
and the Kendall's $\tau$-metric.
In Section~\ref{sec:revCodes} we will review and amend some of the known constructions
of error-correcting codes for permutations and multi-permutations, using the Kendall's
$\tau$-metric. These concepts and constructions will be used in Section~\ref{sec:sysPer} to obtain our main
construction of systematic error-correcting codes for permutations.
We will also perform some analysis for the number
of redundancy symbols of these codes.
We extend this construction to form systematic
error-correcting codes for multi-permutations in Section~\ref{sec:sysMultPer}.
We conclude in Section~\ref{sec:conclusion}.

\section{Basic Concepts}
\label{sec:basic}

We denote by $[n]$ the set of $n$ integers $\{1,2,\ldots,n\}$.
For two integers $a,b$, $a<b$, we denote by
$[a,b]$ the set of $b-a+1$ integers $[a,a+1,a+2,\ldots,b]$.
Let $S_n$ be the set of all permutations on $[n]$,
and let $S([a,b])$ be the set of all permutations
on $[a,b]$. A~more general concept is \emph{multi-permutations}, which is also known as
permutations with repetitions. A~\emph{multi-set}
$\cM=\{v_1^{m_1},v_2^{m_2},\cdots,v_{\ell}^{m_{\ell}}\}$ is a
collection of the elements $\{ v_1 , v_2 , \ldots , v_\ell \}$ in
which~$v_i$ appears $m_i$ times, for each $i$, $1 \leq i \leq \ell$.
The elements of $\{ v_1,v_2,\ldots,v_{\ell} \}$ are called \emph{ranks}
while for every $i$, $1\leq i\leq \ell$,
the positive integer $m_i$ is called
the \emph{multiplicity} of the $i$th rank. If $m_1=m_2 = \cdots=m_{\ell}=m$ then~$\cM$
is called a \emph{balanced multi-set}. A multi-permutation on the multi-set $\cM$
is an ordering of all the elements of~$\cM$.
Note, that a permutation is a special case of a multi-permutation.
We denote a multi-permutation $\sigma$ of length~$n$ by
$\sigma=[\sigma(1),\sigma(2),\ldots,\sigma(n)]$, $n=\sum_{i=1}^{\ell}m_i$.
For example, if $\cM=\{1^{2},2^{3},3\}$, then $\sigma=[1,2,2,1,3,2]$ is a multi-permutation on $\cM$.
We denote by $S(\cM)$ the set of all multi-permutations on $\cM$.
The size of $S(\cM)$ is equal to
$\frac{n !}{\Pi_{i=1}^\ell m_i !}$.

Given a multi-permutation $\sigma=[\sigma(1),\sigma(2),\ldots,\sigma(n)]$ from $S(\cM)$,
an \emph{adjacent transposition} is an exchange of two distinct adjacent elements
$\sigma(i),\sigma(i+1)$, in~$\sigma$, for some $1\leq i\leq n-1$.
The result of such an adjacent transposition is the multi-permutation
${[\sigma(1),\ldots,\sigma(i-1),\sigma(i+1),\sigma(i),\sigma(i+2),\ldots,\sigma(n)]}$.
The Kendall's $\tau$-distance between two multi-permutations
$\sigma,\pi\in S(\cM)$ denoted by $d_K (\sigma,\pi)$
is the minimum number of adjacent transpositions
required to obtain the multi-permutation $\pi$ from the multi-permutation $\sigma$.

\begin{example}
If $\sigma=[1,1,2,2]$ and $\pi=[2,1,2,1]$, then
$d_K(\sigma,\pi)=3$, since at least three adjacent transpositions
are required to change the multi-permutation
$\sigma$ to $\pi$: $[1,1,2,2]\to [1,2,1,2]\to
[2,1,1,2]\to [2,1,2,1]$.
\end{example}

The Kendall's $\tau$-metric was originally defined for permutations \cite{DiGr77,KeGi90}.
For two permutations $\sigma,\pi\in S_n$ it is known~\cite{JSB10,Knu98}
that $d_K(\sigma,\pi)$ can be expressed as\vspace{-0.2cm}

\begin{small}
\begin{equation*}
\label{eq:dkPerms}
d_K(\sigma,\pi)=|\{ (i,j) : \sigma^{-1}(i) < \sigma^{-1}(j),~\pi^{-1}(i) > \pi^{-1}(j) \}|.
\vspace{-0.2cm}
\end{equation*}
\end{small}

\noindent
For a multi-permutation $\sigma\in S(\cM)$, where
$\cM=\{v_1^{m_1},v_2^{m_2},\ldots,v_{\ell}^{m_{\ell}}\}$, we distinguish
between appearances of the same rank in $\sigma$, by their positions
in $\sigma$. We consider the increasing order of these positions. By
abuse of notation we sometimes write $\sigma(j)=v_{i,r}$
and $j=\sigma^{-1} (v_{i,r})$ to indicate
that the $r$th appearance of $v_i$ is in the $j$th
position in $\sigma$. The computation of the Kendall's $\tau$-distance
between two permutations can be generalized to
two multi-permutations $\sigma,\pi\in S(\cM)$ as follows

\begin{small}
\begin{equation*}
d_K(\sigma,\pi)= \left| \left\{((i,r),(j,s))~:~
\begin{array}{c} \sigma^{-1}(v_{i,r}) < \sigma^{-1}(v_{j,s})\\ \pi^{-1}(v_{i,r})>\pi^{-1}(v_{j,s}) \end{array} \right\} \right|.
\end{equation*}
\end{small}

Let $n_0=0$ and for $1\leq i\leq \ell$  let $n_i=\sum_{j=1}^i m_j$,
which implies that $n=n_{\ell}$.
For each $i$, $1\leq i\leq \ell$, let $\theta_i$ be a permutation on
$[ n_{i-1}+1,n_{i}]$ and let ${\bf \theta}=(\theta_1,\theta_2,\ldots,\theta_{\ell})$.
We define a mapping
$T_{\mathbf{\theta}}:S(\cM) \rightarrow S_n,$
such that for every $\sigma \in S(\cM)$, $T_{\mathbf{\theta}}(\sigma)$
is the permutation in~$S_n$ obtained
as follows. For each $i$, $1 \leq i \leq \ell$,
the permutation~$\theta_i$ is substituted, in
the same order, in the~$m_i$ positions in which the rank $v_i$ appears in~$\sigma$.
More precisely, if $\sigma(j)=v_{i,r}$ then $(T_{\mathbf{\theta}}(\sigma))(j)=\theta_i(r)=\theta(r+n_{i-1})$.
For example, if $\mathbf{\theta}=(\theta_1,\theta_2,\theta_3)$, where $\theta_1=[1,3,2]$,
$\theta_{2}=[4,5]$, and $\theta_3=[8,7,6]$, then
$T_{\mathbf{\theta}}([1,2,1,3,3,1,3,2]) = [1,4,3,8,7,2,6,5]).$
The mappings $T_{\mathbf{\theta}}$ are useful for the computation
of the Kendall's $\tau$-distance between two multi-permutations since it
is reduced to the computation of the Kendall's $\tau$-distance
on the corresponding permutations.

\begin{lemma}
\label{lem:TfunctionsDist}
For every two multi-permutations $\sigma,\pi \in S(\cM)$ and
${\bf \theta}=(\theta_1,\theta_2,\ldots,\theta_{\ell})$ we have
$$
d_K (\sigma,\pi) = d_K (T_{\mathbf{\theta}}(\sigma),T_{\mathbf{\theta}}(\pi)).
$$
\end{lemma}

\begin{example}
If $\sigma=[1,1,2,2]$, $\pi=[2,1,2,1]$, $\mathbf{\theta}=(\theta_1,\theta_2)$,
where $\theta_1=[2,1]$ and $\theta_2=[3,4]$, then
$d_K (\sigma,\pi)=3$, and $d_K (T_{\mathbf{\theta}}([1,1,2,2]),T_{\mathbf{\theta}}([2,1,2,1]))=
d_K ([2,1,3,4],[3,2,4,1])=3$.
\end{example}

\begin{lemma}\label{lem:PermutationsGeqMultDist}
Let $\sigma,\pi\in S(\cM)$ and let $\mathbf{\theta}=(\theta_1,\theta_2,\ldots,\theta_{\ell})$,
$\mathbf{\eta}=(\eta_1,\eta_2,\ldots,\eta_{\ell})$,
where $\theta_i,\eta_i\in S([n_{i-1}+1,n_{i}])$, for each $i$, $1\leq i\leq \ell$.
Then
$$
d_K(T_{\mathbf{\theta}}(\sigma),T_{\mathbf{\eta}}(\pi))\geq d_K(\sigma,\pi)+\sum_{i=1}^{\ell}d_K(\theta_i,\eta_i).
$$
\end{lemma}

Another simple and important property of the Kendall's $\tau$-metric
on multi-permutations is presented in the following lemma.

\begin{lemma}\label{lem:biGraph}
If $\sigma$, $\pi$, and $\rho$, are three multi-permutations in $S( \cM )$, then
$d_K (\sigma,\pi) + d_K (\pi,\rho) \equiv d_K (\sigma,\rho) ~(\text{mod}~2)$.
\end{lemma}

\section{Error-Correcting Codes}
\label{sec:revCodes}
For the construction of systematic error-correcting codes for
permutations and multi-permutations
given in Sections~\ref{sec:sysPer} and~\ref{sec:sysMultPer}
we need general error-correcting codes for multi-permutations.
In this section we discuss the constructions for such
error-correcting codes for multi-permutations with the
Kendall's $\tau$-distance.

Such a construction was given in~\cite{SGD12}.
It is based on a metric embedding (mapping) of $S(\cM)$,
where $\cM$ is a balanced multi-set, into the metric space
$\mathbb{Z}^{n-m}$, where $m$ is the multiplicity of
the ranks. The \emph{Manhattan distance} (also
called the $L_1$-distance) is used in $\mathbb{Z}^{n-m}$. This construction is a
generalization of the constructions in~\cite{BM10,JSB10}
for error-correcting codes for permutations.

Let $\mathbf{x},\mathbf{y}\in \mathbb{Z}^{N}$, $\mathbf{x}=(x_1,x_2,\ldots,x_{N})$,
$\mathbf{y}=(y_1,y_2,\ldots,y_N)$. The Manhattan distance
$d_M(\mathbf{x},\mathbf{y})$ is defined by
$$
d_M(\mathbf{x},\mathbf{y}) \deff \sum_{i=1}^{N}|x_{i}-y_{i}|.
$$
This metric embedding (mapping) is injective and for every
two multi-permutations $\sigma$ and $\pi$ in $S(\cM)$,
$d_K (\sigma , \pi )$ is greater or
equal to the Manhattan distance between their images in $\mathbb{Z}^{n-m}$.
These properties allow to construct error-correcting codes in $S({\cM})$
from error-correcting codes in the Manhattan metric over $\mathbb{Z}^{n-m}$.

We present a slightly modified version of this mapping.
It will be defined on $S(\cM)$, where $\cM$ is any multi-set,
not necessarily a balanced multi-set.
We will also restrict its range to its image,
in order to obtain a bijective mapping. This is important for encoding purpose.
We will show an encoding of $S(\cM)$, based on the enumerative
encoding algorithm of Cover~\cite{Cov73}
in the full version of this paper.

A vector $\mathbf{x}=(x_1,x_2,\ldots,x_k) \in \Z^k$ is
\emph{monotone} if $x_1\geq x_2\geq \ldots\geq x_k$.
For a set $S$ of integers let $[S]^k$ be the set of all monotone vectors
of length $k$ over~$S$.
Let
\begin{small}
\begin{equation*}
[\mathbb{Z}]^{\cM} \deff [\mathbb{Z}_{n_1+1}]^{m_2}\times[\mathbb{Z}_{n_{2}+1}]^{m_{3}}\times \ldots\times[\mathbb{Z}_{n_{\ell-1}+1}]^{m_{\ell}}.
\end{equation*}
\end{small}
The mapping $\psi:S(\cM)\rightarrow [\mathbb{Z}]^{\cM}$ is defined as follows.
For every $\sigma\in S(\cM)$, $\psi(\sigma)$ is the vector $\mathbf{x}\in [\mathbb{Z}]^{\cM}$,
$\mathbf{x}=(\mathbf{x}_{2},\mathbf{x}_{3},\ldots, \mathbf{x}_{\ell})$, where for each $i$, $2\leq i\leq \ell$,
$\mathbf{x}_i=(x_{i,1},x_{i,2},\ldots,x_{i,{m_i}})$, and
for each~$s$, $1\leq s\leq m_i$,
$$
x_{i,s} \deff |\{j~:~j>\sigma^{-1}(v_{i,s})\wedge \sigma(j)< i\}|.
$$
Namely, $x_{i,s}$ counts the number of ranks smaller than~$i$ which appear
to the right of the $s$th appearance of $i$.
For example, if $\sigma=[2,1,3,4,3,2,1,4]$
then $\psi(\sigma)=(\mathbf{x}_2,\mathbf{x}_3,\mathbf{x}_4)=((2,1),(2,2),(3,0))$.

\begin{lemma}
\label{lem:psiBijective}
The mapping $\psi$ is bijective.
\end{lemma}

\begin{lemma}
\label{lem:KleqLee}
For any two multi-permutations $\sigma,\pi\in S(\cM)$ we have
$$
d_M(\psi(\sigma),\psi(\pi))\leq d_K(\sigma,\pi).
$$
\end{lemma}

Let $\mathbb{Z}_q^N$ be the set of all vectors of length $N$ over the alphabet $\mathbb{Z}_q$.
For every two vectors $\mathbf{x},\mathbf{y}\in \mathbb{Z}_q^N$,
the Lee distance $d_L(\mathbf{x},\mathbf{y})$ is defined by
$$
d_L(\mathbf{x},\mathbf{y}) \deff \sum_{i=1}^N\min\{|x_i-y_i|,q-|x_i-y_i|\}.
$$

Clearly, $d_M(\mathbf{x},\mathbf{y})\geq d_L(\mathbf{x},\mathbf{y})$ for all
$\mathbf{x},\mathbf{y}\in \mathbb{Z}_q^N$.
The set $[\mathbb{Z}]^{\cM}$ is a subset of $\mathbb{Z}_q^{n-m_1}$,
where $q > n_{\ell-1}$. Hence,
$d_L(\psi(\sigma),\psi(\pi))\leq d_K(\sigma,\pi)$ for every two multi-permutations $\sigma,\pi\in S(\cM)$.
We are now in a position to present a construction which transfers codes with
the Lee metric to codes with the Kendall's $\tau$-metric.
The related theorem
is a slight generalization of the result in~\cite{SGD12}. This construction
will be a major component in our main construction of systematic codes,
which is the primary goal of this paper.

\begin{theorem}
\label{th:mainConstruction}
If there exists a code $\cC_L\subseteq \mathbb{Z}_q^{n-m_1}$, ${q > n_{\ell-1}}$,
with minimum Lee distance $d$,
then there exists a code $\cC_K\subseteq S(\cM)$ with minimum Kendall's
$\tau$-distance at least $d$ and of size $|\cC_K|=|\cC_L\cap [\mathbb{Z}]^{\cM}|$.
\end{theorem}

By Theorem~\ref{th:mainConstruction}, error-correcting codes
in $S(\cM)$ with the Kendall's $\tau$-metric
can be constructed from error-correcting codes over $\mathbb{Z}_q^{n-m_1}$
in the Lee metric.
Next, we present some of the known constructions of error-correcting
codes in the Lee metric and use
Theorem~\ref{th:mainConstruction} to obtain error-correcting codes in $S(\cM)$
and to estimate the size of these codes.
First, we consider single-error-correcting codes in the Lee metric.
Golomb and Welch~\cite{GoWe70} presented the following construction
of a perfect linear single-error-correcting code
in the Lee metric.
\begin{theorem}
\label{thm:GoWe}
For every positive integer $N$, the code
$$
\cC_L=\left\{ \mathbf{x}\in \mathbb{Z}_{2N+1}^N~:~\sum_{i=1}^Ni\cdot x_i\equiv 0~(\hbox{mod }2N+1)\right\}
$$
is a perfect linear single-error-correcting
code  in $\mathbb{Z}_{2N+1}^N$ with the Lee metric.
\end{theorem}

The construction in Theorem~\ref{thm:GoWe}
was used in~\cite{JSB10} to construct single-error-correcting codes for
permutations with the Kendall's $\tau$-distance.
Combining this construction with Theorem~\ref{th:mainConstruction}
implies the following corollary.

\begin{cor}
\label{cor:singleErrorCode}
There exists a single-error-correcting code $\cC_K\subset
S(\cM)$ of size $|\cC_K|\geq \frac{|S(\cM)|}{2(n-m_1)+1}$.
\end{cor}

The following construction was first proposed by Varshamov
and Tenengolts~\cite{VaTe65} (see also~\cite{BM10})
for codes which correct a single asymmetric error.
Let $||\mathbf{x}||$ denote the Manhattan weight of $\mathbf{x}$.

\begin{theorem}
\label{thm:tLee}
Let $q\geq N$ and let $h_1,h_2,\ldots,h_N$ be integers, $0<h_i<q$ for all $1\leq i\leq N$.
Assume that for every~${\mathbf{e}\in \mathbb{Z}^{N}}$ with $||\mathbf{e}||\leq t$,
the sums $\sum_{i=1}^N e_i\cdot h_i$ are all distinct
modulo $q$. Then the code
$$
C=\left\{\mathbf{x}\in \mathbb{Z}_q^N~|~\sum_{i=1}^Nx_i\cdot h_i\equiv 0 ~(\hbox{mod }q)\right\}
$$
is a linear $t$-error-correcting code in~$\mathbb{Z}_q^N$
with the Lee metric.
\end{theorem}

In order to use the construction in Theorem~\ref{thm:tLee}
we need the following theorem of Barg and Mazumdar~\cite{BM10}.

\begin{theorem}
\label{th:getingHi}
Let $q$ be a power of a prime and ${M=(q^{t+1}-1)/(q-1)}$. Let
$$
M_t=\left\{
              \begin{array}{ll}
                t(t+1)M, & t \hbox{ is odd} \\
                t(t+2)M, & t \hbox{ is even}
              \end{array}
            \right.
$$
Then there exist integers $h_1,h_2,\ldots, h_{q+1}$ such that for all
$\mathbf{e}\in \mathbb{Z}^{q+1}$, $||\mathbf{e}||\leq t$,
the sums $\sum_{i=1}^{q+1} e_i h_i$ are all distinct modulo $M_t$.
\end{theorem}

The construction in Theorem~\ref{thm:tLee} of a $t$-error-correcting code
in the Lee metric, combined with Theorem \ref{th:getingHi},
was used in \cite{BM10} to construct $t$-error-correcting codes for permutations with the
Kendall's $\tau$-metric, and also used in \cite{SGD12} to construct $t$-error-correcting
codes with the Kendall's $\tau$-metric for multi-permutations over a balanced multi-set.
Other constructions of codes with the Kendall's $\tau$-distance that might useful
in this context can be found in~\cite{MBZ13}.
By combining the construction in
Theorems~\ref{th:mainConstruction},~\ref{thm:tLee},~and~\ref{th:getingHi}
we obtain the following Corollary.
\begin{cor}
\label{cor:tCodesMult}
Let $M=((n-m_1-1)^{t+1}-1)/(n-m_1-2)$, where $n-m_1-1$ is a power of a prime.
There exists a $t$-error-correcting code $\cC\subset S(\cM)$ in the Kendall's
$\tau$-metric, whose size satisfies
$$
|\cC|\geq \left\{
              \begin{array}{ll}
                \frac{|S(\cM)|}{t(t+1)M}, & t~\hbox{is odd} \\
                \frac{|S(\cM)|}{t(t+2)M}, & t~\hbox{is even}
              \end{array}
            \right.
$$
\end{cor}
\vspace{0.1cm}

Now, after presenting the concepts and ideas in constructions of error-correcting
codes for multi-permutations, we are ready to present our main results
on systematic error-correcting codes for permutations and
multi-permutations in the next two sections.

\section{Systematic ECC for Permutations}
\label{sec:sysPer}
In this section we present systematic $t$-error-correcting codes for permutations.
Let $k,n$ be integers such that $n\geq k \geq 1$. For a permutation $\alpha\in S_n$,
we define $\alpha_{\downarrow k}$ to be the permutation obtained from
$\alpha$ by deleting all the elements of $\{k+1,k+2,\ldots,n\}$ from $\alpha$.
We also define $\alpha_{k\mapsto 0}$ to be the multi-permutation obtained from
$\alpha$ by replacing in $\alpha$ every element of $\{1,2,\ldots,k\}$ by $0$. For example,
if $\alpha=[2,5,4,1,3,6]$ and $k=3$ then $\alpha_{\downarrow k}=[2,1,3]$ and $\alpha_{k\mapsto 0}=[0,5,4,0,0,6]$.
In \cite{ZJB12}, the authors define systematic codes in the following way.
A code $\cC\subseteq S_n$ is an $(n,k)$ \emph{systematic} code if for every ${\sigma\in S_k}$
there exists exactly one ${\alpha\in \cC}$ such that
$\alpha_{\downarrow k}=\sigma$, which implies that
$|\cC|=k!$. The number of \emph{redundancy symbols}
of an $(n,k)$ systematic code is ${r=n-k}$.

Let $r$ be a positive integer and let ${\cM_{k,r} \deff \{0^k,k+1,k+2,\ldots,k+r\}}$.
For every permutation $\sigma\in S_k$ and multi-permutation $\rho\in S(\cM_{k,r})$,
we define the permutation $\sigma \ast \rho$ to be the permutation in
$S_{k+r}$ obtained from $\rho$ by replacing the $k$ zeros in $\rho$ by the $k$ elements
of $\{1,2,\ldots,k\}$, in the same order as in $\sigma$.
For example, if $k=4$, $r=3$, $\rho=[0,6,0,0,5,7,0]$, and $\sigma=[2,4,1,3]$,
then $\sigma \ast \rho =[2,6,4,1,5,7,3]$.
\begin{lemma}
\label{lem:permTomult}
For every $\rho\in S(\cM_{k,r})$ and $\sigma\in S_k$ we have
\begin{itemize}
\item[1)] $(\sigma \ast \rho)_{\downarrow k}=\sigma$.

\item[2)] $(\sigma \ast \rho)_{k \mapsto 0}=\rho$.
\end{itemize}
\end{lemma}

By Lemma~\ref{lem:PermutationsGeqMultDist} we have.
\begin{lemma}
\label{lem:multDist}
Let $\sigma,\pi\in S_k$ and $\rho_1,\rho_2\in S(\cM_{k,r})$. Then
$$
d_K(\sigma \ast \rho_1 ,\pi \ast \rho_2)\geq d_K(\sigma,\pi) + d_K(\rho_1,\rho_2)~.
$$
\end{lemma}

We are now in a position to present our construction for systematic
error-correcting codes for permutations.
\begin{theorem}
\label{th:mainConsSystematic}
Let $h_1,h_2,\ldots,h_{k-1}$, and $Mֹ_t$, be integers such that for every
$\mathbf{e}\in \mathbb{Z}^{k-1}$ with $||\mathbf{e}||\leq t$, the sums
$\sum_{i=1}^{k-1}e_ih_i$ are all distinct modulo $M_t$.
Assume further that there exists a code  $\cC_r\subset S(\cM_{k,r})$ with minimum Kendall's $\tau$-distance $2t$
and of size $|\cC_{r}|\geq M_t$. Let $\rho_0,\rho_1,\ldots,\rho_{M_t-1}$ be distinct multi-permutations
in $\cC_{r}$.
Let $\cC$ be the code in~$S_{k+r}$ defined as follows.
$$
\cC=\{\sigma \ast \rho_j~:~\sigma \in S_k,~\sum_{i=1}^{k-1} (\psi(\sigma))_{i+1} h_i\equiv j ~ \hspace{0cm} mod ~M_t \}.
$$
Then the code $\cC$ is a $(k+r,k)$ systematic $t$-error-correcting code.
\end{theorem}

\begin{example}
\label{ex:con1}
Let $k$ be an integer, let $r=2$, and
let $M_1=2(k-1)+1$. As in Theorem~\ref{thm:GoWe},
for every $\mathbf{e}\in \mathbb{Z}^{k-1}$, $||\mathbf{e}||\leq 1$,
the sums $\sum_{i=1}^{k-1}e_ii$ are all distinct modulo $M_1$.
For the construction, we need a code in $S(\cM_{k,2})$ with minimum
distance 2 and of size at least $M_1$.
To this end, fix a multi-permutation $\rho\in S(\cM_{k,2})$ and consider
the codes $\cC^e_2=\{\gamma\in S(\cM_{k,2})~:~d_K(\rho,\gamma)\equiv 0~(\text{mod}~2)\}$
and $\cC^o_2=\{\gamma\in S(\cM_{k,2})~:~d_K(\rho,\gamma)\equiv 1~(\text{mod}~2)\}$.
By Lemma~\ref{lem:biGraph}, the minimum distance
of both $\cC^e_2$ and $\cC^o_2$ is 2. Clearly, the size of either $\cC^e_2$ or
$\cC^o_2$ is at least $\frac{|S(\cM_{k,2})|}{2}=\frac{(k+2)!}{k!\cdot 2}=\frac{(k+2)(k+1)}{2}$.
For all $k\geq 1$ we have that ${\frac{(k+2)(k+1)}{2}\geq 2(k-1)+1}$ and
hence by Theorem~\ref{th:mainConsSystematic} there exists a
$(k+2,k)$ systematic single-error-correcting code.
\end{example}

\begin{example}
\label{ex:con2}
Let $k$ be an integer such that $k-2$ is a power of
a prime, let $r=3$, and
let $M_2=8((k-2)^3-1)/(k-3)=8((k-2)^2+k-1)$. By Theorem~\ref{th:getingHi}, it follows that there exist
$h_1,h_2,\ldots,h_{k-1}$ such that for all $\mathbf{e}\in \mathbb{Z}^{k-1}$, $||\mathbf{e}||\leq 2$,
the sums $\sum_{i=1}^{k-1}e_ih_i$ are all distinct modulo $M_2$.
We have to show the existence of a code in $S(\cM_{k,3})$ with minimum distance 4 and of size at least $M_2$.
By Corollary~\ref{cor:singleErrorCode}, there exists a single-error-correcting code
$\cC_K\subset S(\cM_{k,3})$ of size $|\cC_K|\geq \frac{|S(\cM_{k,3})|}{2\cdot 3+1}$. We fix a multi-permutation
$\rho\in S(\cM_{k,3})$ and consider the codes $\cC^e_3=\{\gamma\in \cC_K~:~d_K(\rho,\gamma)\equiv 0~(\text{mod}~2)\}$
and $\cC^o_3=\{\gamma\in \cC_K~:~d_K(\rho,\gamma)\equiv 1~(\text{mod}~2)\}$.
By Lemma~\ref{lem:biGraph}, it follows that the minimum distance of the codes
$\cC^e_3$ and $\cC^o_3$ is 4. One of these codes must be of size at least $\frac{|\cC_K|}{2}$. If $\cC_3$ is this code
then $|\cC_{3}|\geq  \frac{|S(\cM_{k,3})|}{14}=\frac{(k+3)!}{k!\cdot 14}=\frac{(k+3)(k+2)(k+1)}{14}$.
For all $k\geq 113$ we have that $\frac{(k+3)(k+2)(k+1)}{14}\geq 8((k-2)^2+k-1)$
and hence by Theorem~\ref{th:mainConsSystematic}, if $k \geq 113$
such that $k-2$ is a power of a prime then there
exists a $(k+3,k)$ systematic double-error-correcting code.
\end{example}

Theorem~\ref{th:getingHi}, Theorem~\ref{th:mainConsSystematic}, and
Corollary~\ref{cor:tCodesMult} lead to the following result
which follows in the same lines as Example~\ref{ex:con1}
and Example~\ref{ex:con2}.

\begin{cor}
Let $t$ be a power of a prime and let $r=t+1$.
Then there exists an integer $K_t$ such that for every integer $k\geq K_t$ for which $k-2$ is a power of
a prime, there exists a $(k+r,k)$ systematic $t$-error-correcting code.
\end{cor}

In~\cite{ZJB12,ZMJB13} a construction of systematic $(k,k+2)$ single-error-correcting codes
for permutations with two redundancy symbols was given.
They have the same number of redundancy symbols as in Example~\ref{ex:con1}.
They construct $(n,k)$ systematic $t$-error-correcting
codes with at most $2t+1$ redundancy symbols. If $k$ and $t$ have the same magnitude
then for some parameters the codes of our construction
have the same number of redundancy symbols,
but for most parameters the number of redundancy symbols of the
codes in our construction is considerably better.
Our main theorem is stated as follows.

\begin{theorem}
\label{thm:main}
Let $k$ be an integer such that $k-2$ is a power of a prime,
let $t=k^{\epsilon}$ be a positive integer, and let $r=\lceil \mu t \rceil$, where $r-1$
is a power of a prime. If $k$ is large enough and if
$$
\left\{
              \begin{array}{lll}
                \mu >1+\epsilon & \mbox{for} & 0\leq \epsilon\leq 1 \\
                \mu >1+\frac{1}{\epsilon} & \mbox{for} & 1<\epsilon~,
              \end{array}
            \right.
$$
then
there exists a $(k+r,k)$ systematic
$t$-error-correcting code.
\end{theorem}
The conditions in Theorem~\ref{thm:main} can be relaxed. If we use
a power of a prime $k'-2$, $k \leq k' \leq 2k$, in Theorem~\ref{th:mainConsSystematic}
and the integers $h_1,h_2,\ldots,h_{k-1}$ from the integers
$h_1,h_2,\ldots,h_{k'-1}$ then we can omit the requirement
that $k-2$ should be a power of a prime. Related arguments
can be used to drop the requirement that $r-1$ is a power of a prime.
The related result is described in Section~\ref{sec:conclusion}.
\section{Systematic ECC for Multi-permutations}
\label{sec:sysMultPer}
In this section we generalize the construction in Section~\ref{sec:sysPer}
to obtain systematic error-correcting codes for multi-permutations.
In the most general definition of systematic
codes for multi-permutations we have a multi-set~$\cK$
with $k$ elements (with repetitions) serving as the
information symbols and a multi-set~$\cR$
with $r$ elements serving as the redundancy symbols. The intersection
between~$\cK$ and~$\cR$ must be empty. The codewords
are multi-permutations over the multi-set
$\cK \cup \cR$. The number of codewords
in the error-correcting code must be the number of distinct
multi-permutations over the multi-set~$\cK$.
In the systematic code $\cC$ each
multi-permutation over the multi-set $\cK$, appears
as a sub-multi-permutation
of exactly one codeword from $\cC$.
The construction
for systematic multi-permutations will be a direct generalization
of the construction in Theorem~\ref{th:mainConsSystematic}.
Instead of the set $\cM_{k,r}$ we use the
set $\cM$ defined by $\cM \deff \{ 0^k \} \cup \cR$,
where 0 is a symbol which does not appear in $\cR$. The size of the code
$\cC_r \subset S(\cM)$ is at least $M_t$ (note, that
the number of $h_i$'s is smaller than $k-1$, unless $\cK$ is a set rather than a multi-set,
and hence $M_t$ will be smaller).

The challenge for systematic permutations codes is to minimize
the number of redundancy symbols of the codes.
For systematic error-correcting codes for multi-permutations there is a tradeoff between the number of
redundancy ranks and the magnitudes of their multiplicities.
For example, in a systematic code for multi-permutations with only
one redundancy rank, the multiplicity of the redundancy rank might be large. However,
by allowing two redundancy ranks, the multiplicity of each redundancy rank should be smaller.
The construction in Theorem~\ref{th:mainConsSystematic}
allows any desirable number of redundancy ranks.

\begin{example}
Let $\cK=\{1^{m_1},2^{m_2},\ldots,\ell^{m_{\ell}}\}$ be a multi-set which consists
of $k=\sum_{i=1}^{\ell}{m_i}$ information symbols, let $\cR=\{\ell+1,\ell+1\}$
and $\cM=\{0^k,\ell+1,\ell+1\}$. Let $M_1=2(k-m_1)+1$.
For every $\mathbf{e}\in \mathbb{Z}^{k-m_1}$, $||\mathbf{e}||\leq 1$,
the sums $\sum_{i=1}^{k-m_1}e_ii$ are all distinct modulo $M_1$.
For the construction, we need a code in $S(\cM)$ with minimum
distance 2 and of size at least $M_1$.
To this end, fix a multi-permutation $\rho\in S(\cM)$ and consider
the codes $\cC^e_2=\{\gamma\in S(\cM)~:~d_K(\rho,\gamma)\equiv 0~(\text{mod}~2)\}$
and $\cC^o_2=\{\gamma\in S(\cM)~:~d_K(\rho,\gamma)\equiv 1~(\text{mod}~2)\}$.
By Lemma~\ref{lem:biGraph} it follows that the minimum distance
of both~$\cC^e_2$ and $\cC^o_2$ is 2. Clearly, the size of either $\cC^e_2$ or
$\cC^o_2$ is at least $\frac{|S(\cM)|}{2}=\frac{(k+2)!}{k!\cdot 2!\cdot 2}=\frac{(k+2)(k+1)}{4}$.
For all $k\geq 1$ we have that $\frac{(k+2)(k+1)}{4}\geq 2(k-m_1)+1$ and
hence by Theorem~\ref{th:mainConsSystematic} there exists a
systematic single-error-correcting code in $S(\cK\cup\cR)$.
\end{example}

\section{Conclusion}
\label{sec:conclusion}

We have considered constructions of systematic error-correcting codes
over permutations and multi-permutations with the Kendall's $\tau$-distance.
The construction is based on error-correcting codes for
multi-permutations. The main result is for
a large enough integer $k$,
a positive integer $t=k^{\epsilon}$, and $r=\lceil \mu t \rceil$.
In this case,
there exists a $(k+r,k)$ systematic
$t$-error-correcting code if
$$
\left\{
              \begin{array}{lll}
                \mu >1+\epsilon & \mbox{for} & 0\leq \epsilon\leq 1 \\
                \mu >1+\frac{1}{\epsilon} & \mbox{for} & 1<\epsilon~.
              \end{array}
            \right.
$$

\vspace{-1ex}
\section*{Acknowledgment}
The work of Sarit Buzaglo and Tuvi Etzion
was supported in part by the U.S.-Israel
Binational Science Foundation, Jerusalem, Israel, under
Grant No. 2012016.
The work of Eitan Yaakobi and Jehoshua Bruck was supported in part by Intellectual Ventures and
an NSF grant CIF-1218005 and in part by the U.S.-Israel
Binational Science Foundation, Jerusalem, Israel, under
Grant No. 2010075. The work of Eitan Yaakobi was done while
he was with the
Electrical Engineering Department, California Institute of Technology,
Pasadena, CA 91125, U.S.A.

\end{document}